\documentclass[
    oneside,
    aps,
    pra,
    superscriptaddress,
    onecolumn,
    a4paper,
    floatfix,
    longbibliography,
    nofootinbib
]{revtex4-2}

\usepackage[american]{babel}
\usepackage[utf8x]{inputenc}

\usepackage{grffile} 

\usepackage{newtxtext,newtxmath}
\usepackage{microtype}

\usepackage{amsmath}

\usepackage{bbm}
\usepackage{bm}
\usepackage{mathtools}
\usepackage{dsfont}
\usepackage{braket}
\usepackage{cancel}
\usepackage{slashed}

\usepackage{graphicx}
\usepackage[svgnames]{xcolor}

\usepackage{siunitx}

\usepackage{subcaption}
\usepackage{ragged2e}
\DeclareCaptionJustification{justified}{\justifying}
\usepackage{geometry}
\usepackage{ragged2e}
\usepackage{array}
\usepackage{tabularx}
\usepackage{booktabs}
\definecolor{cset-aps-blueberry}{RGB}{28,128,158}
\definecolor{cset-aps-blue}{RGB}{46,44,184}
\definecolor{cset-aps-turquoise}{RGB}{0,67,88}
\definecolor{cset-aps-limegreen}{RGB}{190,219,67}
\definecolor{cset-aps-green}{RGB}{31,138,112}
\definecolor{cset-aps-yellow}{RGB}{255,225,25}
\definecolor{cset-aps-orange}{RGB}{253,116,0}
\definecolor{cset-aps-red}{RGB}{219,0,43}

\usepackage{tikz}
\usepackage{pgfplots}
\pgfplotsset{%
    every axis legend/.append style={%
        cells={anchor=west},
        at={(0.96,0.04)},
        anchor=south east,
        font=\scriptsize,
        },
    every axis/.append style={%
        },
    xmajorgrids=true,
    xminorgrids=false,
    minor x tick num=1,
    /pgf/declare function={
        shotnoise(\nO) = (1 + 5*\nO + 6*\nO^2 + sqrt(13 + 43*\nO + 48*\nO^2 + 16*\nO^3))/(2*(4 + 11*\nO + 9*\nO^2));
    },
}
\usetikzlibrary{decorations}
\usetikzlibrary{calc}

\usepackage{pict2e,picture}

\makeatletter
\DeclareRobustCommand{\Arrow}[1][]{%
\check@mathfonts
\if\relax\detokenize{#1}\relax
\settowidth{\dimen@}{$\m@th\rightarrow$}%
\else
\setlength{\dimen@}{#1}%
\fi
\sbox\z@{\usefont{U}{lasy}{m}{n}\symbol{41}}%
\begin{picture}(\dimen@,\ht\z@)
\roundcap
\put(\dimexpr\dimen@-.7\wd\z@,0){\usebox\z@}
\put(0,\fontdimen22\textfont2){\line(1,0){\dimen@}}
\end{picture}%
}
\makeatother

\usepackage{hyperref}
\hypersetup{%
    colorlinks=true,
    linkcolor={cset-aps-red},
    linkbordercolor={cset-aps-red},
    filecolor={cset-aps-orange},
    filebordercolor={cset-aps-orange},
    citecolor={cset-aps-blue},
    citebordercolor={cset-aps-blue},
    urlcolor={cset-aps-green},
    urlbordercolor={cset-aps-green},
    menucolor={cset-aps-limegreen},
    menubordercolor={cset-aps-limegreen},
    breaklinks=true,
    pdfborderstyle={/S/U/W 2},
    pdfpagemode=UseOutlines,
    pdfstartpage={1},
}

\DeclareMathOperator{\cov}{Cov}

\newcommand{\eg}{e.\,g.}
\newcommand{\ie}{i.\,e.}
\newcommand{\cf}{cf.}

\usepackage{wasysym}

\usepackage[clock]{ifsym}
\usepackage{lipsum}

\usepackage{layouts}

\newcommand{\orcid}[1]{\href{https://orcid.org/#1}{\includegraphics[width=7pt]{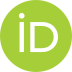}}}

\newcommand{\affTUDa}{Technische Universit{\"a}t Darmstadt, Fachbereich Physik, Institut f{\"u}r Angewandte Physik, Schlo{\ss}gartenstr. 7, D-64289 Darmstadt, Germany}

\newcommand{\affIOF}{Fraunhofer Institute for Applied Optics and Precision Engineering IOF, Albert-Einstein-Str. 7, D-07745 Jena, Germany}

\newcommand{\affHAN}{Institut f{\"u}r Quantenoptik, Leibniz Universit{\"a}t Hannover, Welfengarten 1, D-30167 Hannover, Germany}

\begin{document}

\title{Quantum Imaging Beyond the Standard-Quantum Limit and Phase Distillation}

\author{Simon Schaffrath\,\orcid{0009-0004-4829-6733}\,}
\affiliation{\affTUDa}

\author{Daniel Derr\,\orcid{0000-0002-8690-3897}\,}%
\email{daniel.derr@physik.tu-darmstadt.de, daniel.derr@gmx.net}
\affiliation{\affTUDa}

\author{Markus Gr\"{a}fe\,\orcid{0000-0001-8361-892X}\,}%
\affiliation{\affTUDa}
\affiliation{\affIOF}

\author{Enno Giese\,\orcid{0000-0002-1126-6352}\,}%
\affiliation{\affTUDa}
\affiliation{\affHAN} 

\begin{abstract}
\noindent
Quantum sensing using non-linear interferometers offers the possibility of bicolour imaging, using light that never interacted with the object of interest, and provides a way to achieve phase supersensitivity, \ie{} a Heisenberg-type scaling of the phase uncertainty.
Such a scaling behaviour is extremely susceptible to noise and only arises at specific phases that define the optimal working point of the device.
While phase-shifting algorithms are to some degree robust against the deleterious effects induced by noise they extract an image by tuning the interferometer phase over a broad range, implying an operation beyond the working point.
In our theoretical study, we investigate both the spontaneous and the high-gain regime of operation of a non-linear interferometer.
In fact, in the spontaneous regime using a distillation technique and operating at the working point leads to a qualitatively similar behaviour.
In the high-gain regime, however, typical distillation techniques inherently forbid a scaling better than the standard-quantum limit, as a consequence of the photon statistics of squeezed vacuum.
In contrast, an operation at the working point still may lead to a sensitivity below shot noise, even in the presence of noise.
Therefore, this procedure opens the perspective of bicolour imaging with a better than shot-noise phase uncertainty by working in the vicinity of the working point.
Our results transfer quantum imaging distillation in a noisy environment to the high-gain regime with the ultimate goal of harnessing its full potential by combining bicolour imaging and phase supersensitivity.

\vspace{1mm}
\noindent[This article has been published in \href{https://doi.org/10.1088/1367-2630/ad223f}{New J. Phys. \textbf{26}, 023018 (2024)}.]
\end{abstract}

\maketitle

\vspace{-4mm}
\section{Introduction}
Quantum imaging has emerged as a transformative field, pushing the boundaries of classical optical techniques and offering innovative solutions in diverse applications~\cite{GilaberteBasset2019}. 
At the heart of quantum imaging techniques lies the ability to harness the unique quantum properties of light to achieve unprecedented levels of sensitivity and precision~\cite{Israel2014, Ono2013, Tenne2018, Brida2010, Zander2022, Camphausen2023}. 

Recent developments in quantum imaging have spotlighted the potential for profound advancements, specifically in image distillation, where quantum processes are harnessed to enhance image quality by distillation from undesired noise~\cite{Defienne2019,Tan2008,Lopaeva2013, Fuenzalida2023}.
This can be realised by the use of phase-shifting holography~\cite{Gabor1948, Malacara2007} within the quantum framework, which has unlocked novel possibilities for refining images, bringing a new dimension to the field.
This advancement, when coupled with non-linear interferometers~\cite{Toepfer2022, Haase2022, Black2023, Abramovic2023}, has the potential to redefine the limits of quantum imaging.

Quantum imaging setups comprised of two squeezers (mostly based on non-linear crystals) generating two non-degenerate frequency modes allow for sensing objects at one wavelength, while detecting light at the other wavelength, \ie{} with light that did not interact with the object.
This type of bicolour imaging is called quantum imaging with undetected light~\cite{Lemos2014, Kalashnikov2016} and offers great potential for allowing sensing in spectral ranges, which are normally challenging or impossible to detect.
The working principle of this effect is a missing which-source information, which by itself induces coherence in the low-gain regime~\cite{Wang1991} so that quantum interference can be observed.
Increasing the (parametric) gain beyond the spontaneous regime where no emission is induced still allows for quantum imaging based on induced coherence, even though which-way information may lead to a decreased visibility~\cite{Wiseman2000,Kolobov2017}.

\begin{figure}[t]
    \centering
    \includegraphics[scale=1]{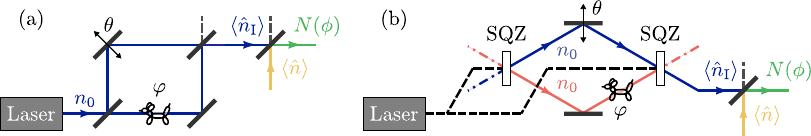}
    \caption{Schematic experimental setup of (a) an MZI and (b) an NLI.
    (a) In an MZI the input laser beam of intensity $n_0$ is split by a first BS into two arms.
    The phase difference $\theta$ between both arms can be tuned by moving the mirror on the upper arm, while the phase object of interest is placed in the lower arm, which in turn imprints the unknown phase $\varphi$ on the lower arm.
    Both arms interfere on a second BS, and the output interferometer signal is described by $\langle \hat{n}_\mathrm{I} \rangle$.
    To model a noisy interferometer signal, this output is incoherently superimposed with noise $\langle \hat{n} \rangle$ on another BS with transmittance $\eta$.
    Finally, the noisy signal $N(\phi)$ is detected which still depends on the total interferometer phase $\phi = \varphi + \theta$.
    (b) In an NLI, a laser seeds two non-degenerate squeezers (SQZs) with two output modes, possibly of different wavelengths (orange and blue).
    The first squeezer has no additional input, so that the number of generated photons $n_0$ in both signal and idler arms solely depends on the gain, which can be tuned from the spontaneous (low-gain) to the high-gain regime.
    The signal photons travel along the upper arm where one can tune the phase difference $\theta$, while the idler photons on the lower arm probe a phase object, which imprints its phase $\varphi$.
    Both modes are redirected to seed the second squeezer, which can be operated at possibly different gain.
    The output in the idler mode $\langle \hat{n}_\mathrm{I} \rangle$, which never interacted with the object, contains all the phase information.
    Again, we model the noisy signal by incoherently  superimposing the output with noise, giving rise to the detected signal $N(\phi)$.}
    \label{fig:interferometer_setups}
\end{figure}

In fact, symmetric setups as shown in figure~\ref{fig:interferometer_setups}\,(b) where both output modes of the first squeezer (SQZ) are used as an input to the second one do not suffer from these subtleties and have led to a class of non-linear interferometers (NLIs)~\cite{Chekhova2016} that are based~\cite{Yurke1986} on the Lie group SU(1,1).
Their study has been met with growing interest in the last few years~\cite{Kviatkovsky2020, RojasSantana2022, GilaberteBasset2021, Panahiyan2023, GilaberteBasset2023}.
Besides the possibility for bicolour imaging with undetected light, one of the key features of such interferometers is---in the ideal case---a phase sensitivity at the Heisenberg limit~\cite{Yurke1986, Chekhova2016}.
However, a sensitivity below  the standard-quantum limit (shot noise) is compromised by internal or detection loss~\cite{Marino2012, Linnemann2016}, which can only partially be compensated by unbalancing the gain in both squeezers~\cite{Manceau2017, Giese2017}.
Hence, loss and noise limit the range of phases where a phase supersensitivity can be achieved.

Our contribution to this burgeoning field is to combine the concept of image distillation via phase-shifting holography within non-linear interferometry.
Since image distillation requires a phase scan that might lie outside the range where phase supersensitivity is observed, we focus on the implications for phase sensitivity.
We discuss the impact of noise on signals in both the low-gain (spontaneous) and high-gain regime for phase imaging.

Such imaging techniques are inherently multi-mode, due to (i) the spatial extent of the object and (ii) the intrinsic multi-mode nature of the radiation generated during (high-gain) parametric down-conversion~\cite{Lemieux2019}.
Assuming that a monochromatic plane wave as a pump field induces the squeezing process, each plane-wave mode of the generated radiation can be treated as an independent SQZ~\cite{Scharwald2023}.
If the object is placed in the far field of the first SQZ, each one of these modes can be assigned to one point of the object, which is subsequently imaged onto the second SQZ~\cite{Frascella2019}.
A detection of the interference signal in the far field of the second SQZ with a camera can then be used to uniquely associate each pixel with a point of the object.
For such a scenario a single-mode description is sufficient, which we present below.
However, we give a more detailed perspective towards the multi-mode scenario and discuss further details in the conclusions.

Since it is a central part of our work, we model noisy signals by incoherently overlaying noise $\braket{\hat{n}}$ onto the signal of an interferometer $\braket{\hat{n}_\mathrm{I}}$ as depicted in figure~\ref{fig:interferometer_setups}.
The total detected phase $\phi = \varphi + \theta$ can be scanned by varying $\theta$ via a movable mirror, such that an object with its phase $\varphi$ can be sensed.
We compare our findings with the performance of a Mach-Zehnder interferometer (MZI) as a bench-marking system, inherently limited by shot noise.
As figure~\ref{fig:interferometer_setups}\,(a) shows, in the MZI, a laser at intensity $n_0$ serves as the interferometer input.
This input is split into two arms, which then interfere with each other after probing the object.
Along these lines, figure~\ref{fig:interferometer_setups}\,(b) shows the case of an NLI.
Here, two squeezers are coherently pumped by a laser.
The first squeezer generates two modes (possibly of different colour) each with intensity $n_0$ through parametric down-conversion.
While the interferometer phase $\theta$ is scanned in one mode, the object with phase $\varphi$ is sensed by the other mode.
Both modes are used as an input to the second squeezer, resulting in a phase-sensitive amplification or absorption.

\section{Interference Signal Superimposed With Noise}
\label{sec:signal_noise}
For the single-mode description of a noisy interference signal, we assume that the information of the imaged object, \ie{} its phase $\varphi$ and transmittance $T$, is completely contained in the output $\langle \hat{n}_\mathrm{I} \rangle$ of the interferometer.
To describe imaging of a spatially extended object, one could assign an index to all quantities in the following calculations, representing the signal detected by one pixel of the camera, which is connected to one plane-wave mode of radiation and one point of the object.
For such a description, all modes of radiation have to be squeezed independently.
In the following we omit such an index, but keep in mind that a multi-mode generalisation is possible under certain assumptions.

\begin{figure}[!htb]
    \centering
    \includegraphics[scale=1]{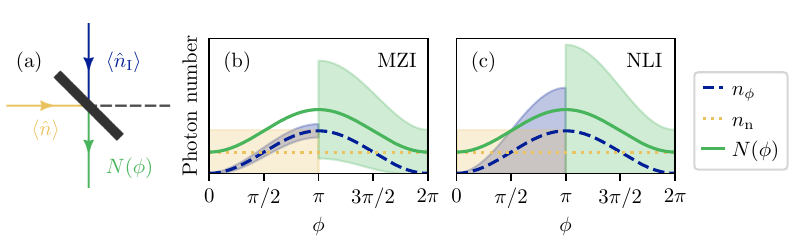}
    \caption{Model for the generation of noisy signals and their resulting statistics.
    (a) An interferometer signal $\langle \hat{n}_\mathrm{I} \rangle$ is incoherently superimposed with noise $\langle \hat{n} \rangle$ on a BS of transmittance $\eta$.
    Afterwards the noisy interferometer signal $N(\phi)$ is detected, which depends on the total interferometer phase $\phi=\varphi + \theta$ that is tunable.
    (b) A fringe scan for an exemplary noisy interferometer signal (solid green line) for Poissonian statistics of the interferometer and thermal statistics of the noise.
    We also show the individual contributions after the BS, noise $n_\text{n}=(1-\eta)\langle \hat{n} \rangle$ (dotted yellow line) and interferometer $n_\phi=\eta \langle \hat{n}_\mathrm{I} \rangle$ (dashed blue line), and illustrate their respective uncertainties (standard deviation) by shaded regions (due to their symmetry with respect to $\pi$, we draw $\Delta n_\phi$ and $\Delta n_\text{n}$ for $\phi<\pi$, as well as $\Delta N(\phi)$ for $\phi>\pi$).
    A Poissonian interferometer output arises for example in an MZI with a laser as input.
    (c) Contributions to the noisy interference signal, where now the respective uncertainties are given for thermal statistics of the interferometer output and noise.
    Since squeezed vacuum has a thermal photon statistics, this case corresponds to the observed signal in quantum imaging.}
    \label{fig:signal_definitions}
\end{figure}

To model the influence of noise, we incoherently superimpose this perfect signal with an additional noise source $\langle \hat{n} \rangle$ on a beam splitter (BS), see figure \ref{fig:signal_definitions}\,(a).
Here, we introduced the operators $\hat{n}_\mathrm{I}$ and $\hat{n}$ describing the photon numbers of one exit port of the interferometer and a corresponding noise mode, respectively.
In the usual fashion, we describe the superposition of two modes on a BS or squeezer by linear transformations~\cite{Giese2017}.
Consequently, the detected signal
\begin{equation}
    N(\phi) =  n_\phi + n_\mathrm{n}
\end{equation}
consists of two contributions:
(i) The transmitted fraction of the interferometer output $n_\phi = \eta \langle \hat{n}_\mathrm{I} \rangle$ for a BS with transmittance $\eta$.
Likewise, as the interferometer output, its detected fraction again contains all of the object's information, in particular the interferometer phase $\phi$.
(ii) The fraction of  noise $n_\mathrm{n} = (1-\eta) \langle \hat{n} \rangle$ coupling into the detector.
Since typically the signal and noise have no common source, we can rule out any interference between the interferometer output and the noise source.
Figure~\ref{fig:signal_definitions}\,(b) and (c) show typical detected interference patterns with individual contributions of interferometer output, noise, and the detected signal for Poissonian and thermal interferometer statistics, respectively.
The latter is of relevance for quantum imaging, since it resembles the statistics of squeezed vacuum.
Exactly this noise model has already been studied experimentally~\cite{Fuenzalida2023} in quantum imaging distillation, even though one can also develop a description where noise couples into the interferometer itself.
Moreover, detection loss can be added trivially by adding another BS before detection with vacuum input~\cite{Giese2017}.

As we observe in figure \ref{fig:signal_definitions}\,(b) and (c), noise not only reduces the detected fraction of the interferometer output, but also affects its uncertainty.
We quantify the uncertainty of the detected signal by its variance and find
\begin{equation}
\label{eq:variance_of_detected_signal}
   \Delta N^2(\phi) = \Delta  n_\phi^2 + \Delta n^2_\mathrm{n} + 2 n_\phi n_\mathrm{n} + (1-\eta) n_\phi + \eta n_\mathrm{n} + 4 \cov(n_\phi, n_\mathrm{n}).
\end{equation}
Apparently, it consists of more contributions than the sum of the individual uncertainties, \ie{} the variance of the detected interferometer signal $\Delta  n_\phi^2 = \eta^2 \langle \hat{n}_\mathrm{I}^2 \rangle- n_\phi^2$, and the variance of the detected noise $\Delta  n_\mathrm{n}^2 = (1-\eta)^2 \langle \hat{n}^2 \rangle- n_\mathrm{n}^2$.
In addition, the covariance $\cov(n_\phi, n_\mathrm{n}) = \eta (1-\eta)\langle \hat{n}_\mathrm{I}  \hat{n} \rangle - n_\phi n_\mathrm{n}$ and further terms enter.
These additional terms are caused by the bosonic commutation relations of the photon creation and annihilation operators used to describe the input modes of the BS that is modelled by an SU(2) transformation, which has a direct impact on the observed photon statistics.
In fact, even when considering vanishing noise, \ie{} $n_\text{n}=0$, the additional term $(1-\eta)n_\phi$ survives and modifies the detected statistics.
This effect can be interpreted as a direct consequence of vacuum fluctuations coupling into the signal and can also be observed for photon detectors with imperfect detection efficiency~\cite{Giese2017}.
We provide further details on the mathematical origin in appendix \ref{sect:origion_of_uncertainty}.

In the following, we assume a vanishing covariance (since the interferometer output and noise are uncorrelated) and Poissonian or thermal statistics for both BS inputs, \ie{} $\Delta \hat{n}_i^2 = \langle \hat{n}_i \rangle$ or $\Delta \hat{n}_i^2 = \langle \hat{n}_i \rangle \langle \hat{n}_i +1 \rangle $, respectively.
With these assumptions and for low interferometer output as well as noise, \ie{} $\langle \hat{n}_\mathrm{I} \rangle, \langle \hat{n} \rangle \ll 1$, the variance reduces to $\Delta N^2(\phi) = n_\phi + n_\mathrm{n}$, independent of the statistics.

\section{Phase Distillation}
\label{sec:phase_distl}
\begin{figure}[h!]
    \centering
    \includegraphics[scale=1]{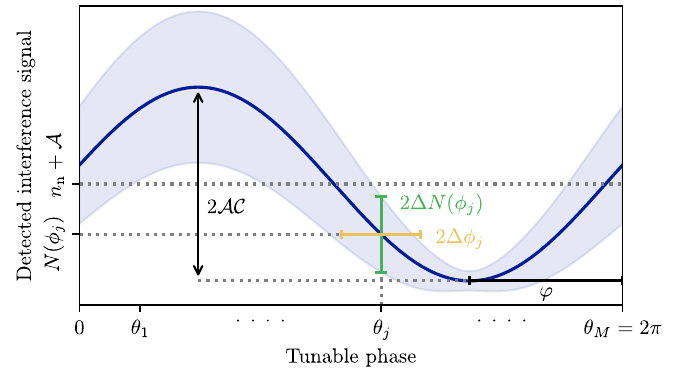}
    \caption{Phase scan of a noisy interferometer signal $N(\phi) = N(\varphi + \theta)$ for quantum imaging distillation.
    We plot the detected interference signal (solid line) against the tunable phase $\theta$.
    A typical fringe has an amplitude $\mathcal{A}$, contrast $\mathcal{C}$, and offset phase $\varphi$ to be determined by the experiment.
    In our model the fringe oscillates around $n_\text{n} + \mathcal{A}$, where $n_\text{n}$ is the detected fraction of noise.
    The distillation technique consists of $M$ measurements of the noisy signal  $N(\phi_j)$ at different phases, \ie{} $\phi_j = \varphi + \theta_j$, and is associated with an uncertainty $\Delta N(\phi_j)$, which in turn gives rise to a phase uncertainty $\Delta \phi_j$ highlighted in the figure.}
    \label{fig:phase_sensitivity}
\end{figure}

In this section, we present a method to extract and reconstruct the phase $\varphi$ of the object from the detected noisy signal.
To obtain an explicit expression, we have to specify the way the information on the object is contained in the detected signal.
For an interference signal, the general form
\begin{equation}
\label{eq:n_phi}
    n_{\phi} =\mathcal{A} \left[ 1 - \mathcal{C} \cos{\phi} \right]
\end{equation}
of the detected fraction of the interferometer output (without noise) is determined by an amplitude $\mathcal{A}$ that already includes the transmittance $\eta$ of the BS, a contrast $\mathcal{C}$, and the total interferometer phase $\phi$.
The contrast contains information on the transmittance of the object, the setup of the interferometer, which might include unbalanced BSs or gains of the squeezers, as well as internal loss.
While the latter is not modelled in our article, its effects can be included straightforwardly, \eg{} by introducing additional BSs.
The phase $\varphi$ of the object is contained in the total interferometer phase via
\begin{equation}
    \phi = \varphi + \theta,
\end{equation}
where $\theta$ is the tunable phase controlled by the experimenter and plays a key role for the distillation procedure.
In fact one collects data for at least three different values of $\theta$, see figure \ref{fig:phase_sensitivity}, which is sufficient to distil the object's phase.

While there exists a variety of algorithms to infer the phase by weighting and combining the collected data~\cite{Malacara2007}, we focus on the procedure with a set of $M\ge 3$ measurements of equally spaced phases $\theta_j \coloneqq j 2 \pi /M$, with $j= 1,2,...,M$ see figure \ref{fig:phase_sensitivity}.
In this case and for a signal of the form of \eqref{eq:n_phi}, the phase information of the object can be reconstructed by weighing the measured results for the individual settings by trigonometric functions, and one finds
\begin{equation}
    \varphi= -\arctan{\left( \frac{\sum_{j=1}^M N(\phi_j) \sin{\theta_j}}{\sum_{j=1}^M N(\phi_j) \cos{\theta_j}} \right)} \quad \mbox{with} \quad \phi_j \coloneqq \varphi + \theta_j .
\end{equation}

To quantify the quality of this distillation technique, we analyse the phase uncertainty through Gaussian error propagation
\begin{equation}
\label{eq:dist_procd_uncert}
    \Delta \varphi^2 = \sum_{j=1}^M \left[ \left( \frac{\partial \varphi}{\partial N(\phi_j)} \right)^2 \Delta N^2(\phi_j) + \left( \frac{\partial \varphi}{\partial \theta_j} \right)^2 \Delta \theta_j ^2 \right] \eqqcolon \Delta \varphi_N^2 + \Delta \varphi_\theta^2,
\end{equation}
which consists of two contributions: (i) The inherent phase uncertainty due to the algorithm $\Delta \varphi_N^2$ depends on the intensity fluctuations $\Delta N^2(\phi_j)$ of the individual measurements.
(ii) The scanning uncertainty $\Delta \varphi_\theta^2$ is caused by the uncertainty of the experimentally tunable phase $\theta$.
In the following, we assume that the first term dominates, while we discuss some requirements on the precision $\Delta \theta_j^2$ of the tunable phase $\theta_j$ in appendix \ref{sect:uncertainty_of_tunable_phase}.
Besides, the impact of errors due to an inaccurate adjustment of $\theta$ on phase-shifting algorithms has already been discussed~\cite{Ayubi2016}.
With the help of the relation $\partial \varphi/\partial N(\phi_j) = 2\sin{ \phi_j }/( M \mathcal{A} \mathcal{C})$ and using trigonometric addition theorems, we arrive at the phase uncertainty 
\begin{equation}
    \label{eq:phase_uncert_dist}
    \Delta \varphi_N^2 = \frac{4\Delta \varphi_0^2}{M \mathcal{C}^2} + \frac{1 + \mathcal{C}^2/4}{M \mathcal{C}^2} 2 \lambda_\phi 
\end{equation}
of the distillation technique, where we have defined the intrinsic phase uncertainty
\begin{equation}
\label{eq:varphi_0}
    \Delta \varphi_0^2 \coloneqq \frac{1}{2} \bigg[  \frac{1+\xi}{\mathcal{A}} + 2 \xi +  \xi^2 \lambda_\mathrm{n}\bigg].
\end{equation}
Here, the parameter $\xi \coloneqq n_\mathrm{n}/\mathcal{A}$ is the ratio of the introduced noise and the amplitude of the interference signal.

To keep a general expression, we allowed for different statistics of the interferometer output encoded in $\lambda_\phi$, while the statistics of the introduced noise is encoded in $\lambda_\text{n}$.
For Poissonian statistics these parameters vanish, while for thermal statistics they are unity.
With this notation, we find the relation $\Delta n_\phi ^2+(1-\eta) n_\phi = n_\phi (1+n_\phi\lambda_\phi )$ as well as $\Delta n_\mathrm{n} ^2+\eta n_\mathrm{n} = n_\mathrm{n} (1+n_\mathrm{n}\lambda_\mathrm{n} )$.

Without noise, we observe from \eqref{eq:varphi_0} that the intrinsic phase uncertainty $\Delta \varphi_0 ^2$ scales as $1/\mathcal{A}$.
Moreover, it is suppressed by the number of measurements $M$, as one would expect from independent measurements.
In section~\ref{sec:MZvsNLI} we demonstrate that  $\Delta \varphi_0 ^2$ exhibits a shot-noise scaling for an MZI operated with coherent light, while for an NLI we observe Heisenberg scaling.
Moreover, this scaling behaviour is modified and masked in a noisy environment, degrading the quality of the distillation procedure.  

In addition to this intrinsic behaviour, a second term contributes to \eqref{eq:phase_uncert_dist} whenever the interferometer displays a thermal photon-number statistics with $\lambda_\phi =1$.
Due to the properties of the generated squeezed light, this case applies in particular to an NLI. 
In contrast to $\Delta \varphi_0 ^2$, this contribution does not depend on the amplitude $\mathcal{A}$ of the interferometer signal and therefore cannot be suppressed by increasing the intensity.
It also arises for perfect contrast and even without any superimposed noise, so that its effect can only be reduced by increasing the number $M$ of measuring points.

In fact, this contribution is inherent to the distillation technique and a consequence of the fact that one averages over measurements at different $\phi_j$, each of which is connected to an uncertainty $\Delta \phi_j^2$.
Using $\partial \varphi/\partial N(\phi_j) = 2\sin{ \phi_j }/( M \mathcal{A} \mathcal{C})$ in the first step, we can therefore represent \eqref{eq:phase_uncert_dist} by
\begin{equation}
\label{eq:distl_connection_to_avg}
    \Delta \varphi_N^2  =\frac{4}{M^2 \mathcal{A}^2 \mathcal{C}^2} \sum_{j=1}^M \sin^2\phi_j \left( \frac{\partial N(\phi_j)}{\partial \phi_j} \right)^2 \, \Delta \phi_j^2%
    = \frac{4}{M^2}\sum_{j=1}^M \sin^4\phi_j \, \Delta \phi_j^2.
\end{equation}
For an NLI, the uncertainty $\Delta \phi_j^2$ strongly depends on the value of $\phi_j$ itself, so that the average leads to a mediocre sensitivity.
However, since NLIs are known to exhibit Heisenberg scaling, this effect can be avoided by solely measuring at a phase setting that is chosen optimally.

\section{Working Point}
As explained above, the distillation technique estimates $\varphi$ by performing measurements at different phases $\phi_j$ that are set by $\theta_j$.
Because each phase setting is associated with its individual uncertainty $\Delta \phi_j^2$ that may vary strongly between different measurements, averaging will not necessarily be optimal.
We therefore minimise the phase uncertainty for a single measuring point $\phi_j$ in this section.

Recalling that the phase of the object is inferred from $\varphi = \phi_j - \theta_j$, the uncertainty observed at one phase setting within one measurement takes the form
\begin{equation}
\label{eq:varphi_uncert}
    \Delta \varphi_j^2  = \Delta \phi_j^2 + \Delta \theta_j ^2 = \frac{\Delta N^2(\phi_j)}{ |\partial_{\phi_j} N(\phi_j)|^2 } + \Delta \theta_j ^2.
\end{equation}
Similar to the distillation technique, we assume that $\Delta \theta_j^2$ is negligible compared to $\Delta \phi_j^2$, see appendix~\ref{sect:uncertainty_of_tunable_phase} for further details.
Minimising $ \Delta \phi_j^2$, we identify a phase $\theta_\text{WP}$ at which the interferometer is ideally operated, referred to as the (optimal) \emph{working point} (WP).
For an interference signal of the form of \eqref{eq:n_phi}, superimposed with noise and measured $M$ times at the same setting, such an optimisation gives rise to the minimal phase uncertainty
\begin{subequations}
\label{eq:phase_uncert_wp_all}
\begin{equation}
    \label{eq:phase_uncert_wp}
    \Delta \phi_\mathrm{WP}^2 = \frac{\Delta \varphi_0^2}{M \mathcal{C}^2}
    + \frac{1 - \mathcal{C}^2  }{2 M \mathcal{C}^2}\lambda_\phi
    + \frac{\sqrt{\varphi_1}}{2M\mathcal{C}^2},
\end{equation}
where we have introduced for compactness the notation
\begin{equation}
    \label{eq:phi_1}
    \varphi_1 \coloneqq%
    \xi \left( \frac{1}{\mathcal{A}} + \xi\lambda_\mathrm{n}  \right) \left[ 4 \Delta \varphi _0 ^2 + 4 \lambda_\phi - \xi \left( \frac{1}{\mathcal{A}} + \xi \lambda_\mathrm{n}\right) \right]%
    + (1-\mathcal{C}^2) \left[\left(\frac{1}{\mathcal{A}} +%
    2\xi + 2\lambda_\phi \right)^2-\lambda_\phi \left( 4 \Delta \varphi_0^2 + \mathcal{C}^2+3 \right)\right].
\end{equation}
\end{subequations}
The phase uncertainty $\Delta \phi_\mathrm{WP}^2$ observed at the WP consists of three contributions:
(i) The first one is again given by the intrinsic phase uncertainty $\Delta \varphi_0^2$ and four times smaller compared to the corresponding contribution in the presented distillation technique.
It therefore results in the same shot-noise or Heisenberg scaling for different types of interferometers, as detailed below.
(ii) For an interferometer signal with thermal statistics, this behaviour is deteriorated by the second term that is not suppressed by intensity and independent of the superimposed noise.
While such an effect also arises in the distillation technique, see \eqref{eq:phase_uncert_dist}, it vanishes for perfect contrast.
This feature is a manifestation of the fact NLIs are extremely susceptible to loss.
(iii) The third contribution to \eqref{eq:phase_uncert_wp} has no counterpart in the distillation procedure.
It is determined by \eqref{eq:phi_1} and vanishes in a setup without noise and for perfect contrast, highlighting once more that increasing noise and imperfect contrast both deteriorate the phase uncertainty.

Overall, the phase uncertainty $\Delta \phi_\mathrm{WP}^2$ at the WP benefits from a perfect setup making it almost independent of the photon statistics of the interferometer signal.
Only this feature makes an operation below shot-noise possible.
However, some of this advantage is lost in presence of noise, which is one of the main reasons to apply the distillation procedure.

\section{Mach-Zehnder Versus Non-linear Interferometer}
\label{sec:MZvsNLI}
To study the metrological benefit of quantum-optically manipulated light and its viability in the distillation procedure, we compare two types of interferometers: an MZI, where the object is interrogated by a coherent state, and an NLI without seed, where the object is interrogated by a squeezed state.

Since the input to the MZI depicted in figure~\ref{fig:interferometer_setups}\,(a) is a coherent state with photon number $n_0$, the overall statistics of the interference signal is also Poissonian, \ie{} we choose $\lambda_\phi = 0$.
A first BS with transmittance $T_1$ generates two arms, where one of them passes through the object that imprints the phase $\varphi$ to be determined by the measurement.
The phase difference of both arms can be tuned by adjusting the path length of the other arm, encoded in $\theta$, which is key to the distillation technique.
Both arms then interfere on a second BS with transmittance $T_2$, whose output is superimposed with noise as described in section~\ref{sec:signal_noise}.
Without internal losses, the amplitude $\mathcal{A} = \eta \gamma n_0$ of the signal scales with the input photon number $n_0$, where we have introduced the factor $\gamma \coloneqq T_1 T_2 + (1-T_1)(1-T_2)$ that describes the imbalance of both BSs. 
The contrast $\mathcal{C} = 2 \sqrt{T_1 T_2 (1-T_1)(1-T_2)}/\gamma$ of the signal reduces to unity for $T_1=1/2=T_2$.
The scaling of the amplitude with $n_0$ is a direct signature of shot noise, as we show below.

For an NLI, both BSs are replaced by non-linear optical elements that generate squeezed light, see figure~\ref{fig:interferometer_setups}\,(b).
In the following discussion, we assume that the NLI is not seeded, so that the first squeezer generates two-mode squeezed vacuum induced by a pump beam, even though a generalisation to squeezed coherent states is in principle possible.
After the non-linear process, each output mode of the squeezer that represents an arm of the NLI carries $n_0$ photons.
The output modes may be non-degenerate in frequency, which allows for quantum and bicolour imaging in both the low- ($n_0\ll 1$) and high-gain ($n_0\gg 1$) regimes of squeezing.
After interacting with a (pure phase) object in one mode, \ie{} assuming no internal losses, both modes are used as a seed of a second squeezer, that either squeezes the state further or leads to anti-squeezing, depending on the phase difference between both arms that can be tuned by $\theta$.
The gain of this squeezer can be identified with a theoretical photon number $n_0'$ that is generated without any seed.
Because the output of the second squeezer is again a two-mode squeezed vacuum, it obeys a thermal photon statistics, \ie{} we choose $\lambda_\phi = 1$.
The amplitude $\mathcal{A}= \eta (n_0 + n_0' + 2n_0 n_0')$ of the interference signal depends on both gains $n_0$ and $n_0'$, and its contrast takes the form $\mathcal{C} = 2 \eta \sqrt{n_0 n_0' (n_0 + 1) (n_0' + 1)}/\mathcal{A}$~\cite{Giese2017}.
For equal gain, that is for $n_0=n_0'$, we observe the signature of Heisenberg scaling since the amplitude becomes proportional to $n_0^2$ for $n_0 \gg 1$.

The contrast, the amplitude, as well as the intensity fluctuations of both MZI and NLI can be further modified, \eg{} by internal and detection loss, an imperfect transmittance of the object, or by mode mixing during the squeezing process itself.
While such effects can be included in our treatment, we refrain from this discussion to focus on the essence of the distillation technique.
For that, we focus in the following on three cases: (i)~imperfect contrast without noise, (ii)~perfect contrast with superimposed noise, and (iii)~the spontaneous regime.

\subsection{Imperfect Contrast Without Noise}
For a first comparison, we discuss the effects of imperfect contrast, \ie{} $\mathcal{C}<1$, to highlight the intrinsic differences between both setups.
Hence, we assume that the signal is not superimposed with noise, \ie{} we choose $\eta=1$ and $\xi=0$.
Since our model includes no internal loss or similar mechanisms, the only way to reduce contrast is to use unbalanced BSs (MZI) or different gains in the squeezers (NLI), respectively.
However, a reduction in contrast may be caused by other effects, such that our results may be generalised to other situations.
The phase uncertainties of the distillation procedure and at the WP for vanishing noise and arbitrary contrast are summarised in table~\ref{tab:results_no_noise} as a special case of \eqref{eq:phase_uncert_dist} and \eqref{eq:phase_uncert_wp_all}.

\begin{table}[htb]
    \caption{Comparison of phase uncertainties for vanishing noise between an MZI and NLI, as well as between the distillation technique $\displaystyle \Delta \varphi_N^2$ and an operation at the WP $\displaystyle \Delta \phi_\text{WP}^2$.
    For both interferometer types, we assume $M$ measuring points and a contrast $\mathcal{C}$.
    In the MZI, $n_0$ describes the intensity of the incoming laser beam and $\gamma$ the imbalance of both BSs, while in the NLI we have two gains $n_0$ and $n_0'$ of the squeezers, which can be interpreted as a hypothetical photon number generated without any seed.
    While we always observe for the MZI shot-noise scaling, the sensitivity of the distillation technique in the NLI is limited by a gain-independent term.
    However, an operation at the WP only gives rise to a Heisenberg-type scaling for perfect contrast.
    }
\label{tab:results_no_noise}
\begin{center}
    \begin{tabular}{lcc}
    \toprule
    & MZI & NLI  \\
    \midrule
    $\displaystyle \Delta \varphi_N^2$ &%
    $\displaystyle \frac{2}{M \mathcal{C}^2 \gamma n_0}$ &%
    $\displaystyle \frac{2}{M \mathcal{C}^2 (n_0 + n_0' + 2n_0 n_0')} + \frac{2+\mathcal{C}^2/2}{M \mathcal{C}^2}$ \\[2em]
    $\displaystyle \Delta \phi_\text{WP}^2$ &%
    $\displaystyle \frac{1}{2 M \gamma n_0 \left(1 - \sqrt{1 - \mathcal{C}^2}\right)}$ &%
    $\quad \displaystyle \frac{1}{2 M \mathcal{C}^2} \left( \frac{1}{n_0 + n_0' + 2n_0 n_0'} + (1 - \mathcal{C}^2) + \sqrt{(1 - \mathcal{C}^2) \left[ \left(\frac{1 + n_0 + n_0' + 2n_0 n_0'}{n_0 + n_0' + 2n_0 n_0'} \right)^2 - \mathcal{C}^2 \right]} \right)$ \\
    \bottomrule
    \end{tabular}
\end{center}
\end{table}

For perfect contrast $\mathcal{C}=1$, implying $\gamma = 1/2$ and $n_0=n_0'$, we already gain two general insights:
(i) Operating an MZI at the WP has a phase uncertainty of a factor $1/4$ lower than found for the distillation procedure, yet both modes of operation are shot-noise limited and show a scaling with $1/n_0$ of the variance.
(ii) In contrast, operating an NLI at the WP leads to a Heisenberg scaling, \ie{} a scaling with $1/n_0^2$ of the variance.
However, the distillation procedure suffers from including less-than-optimal measuring points as discussed in \eqref{eq:distl_connection_to_avg}.
Therefore, it is limited by the second term whose contribution can only be reduced by more measuring points and is not suppressed by increasing the gain.

For imperfect contrast, \ie{} $\mathcal{C}<1$ the phase uncertainty increases in all configurations.
In an MZI the difference between the imaging distillation procedure and measuring at the WP manifests only in a numerical advantage of the WP, while the scaling with intensity is not affected.
The reason lies in the Poissonian output statistics of the MZI \cf{} \eqref{eq:phase_uncert_dist} and \eqref{eq:phase_uncert_wp_all}, where $\lambda_\phi =0$.
However, the NLI is more drastically affected by imperfect contrast.
In the high-gain regime, \ie{} $1 \ll n_0$, the distillation procedure has a major drawback due to the thermal interferometer output statistics.
Only the first term of $\Delta \varphi_N^2$ shows a Heisenberg scaling, while in fact the uncertainty is dominated by the second term independent of any gain and arises for both perfect and imperfect contrast.
This contribution can only be reduced by increasing the number of measuring points.
While in a perfect configuration, operating at the WP leads to a Heisenberg scaling of the uncertainty, this advantage gets diminished by imperfect contrast.
Similar to the distillation procedure, additional terms arise that can only be reduced by increasing $M$ and are independent of the gain or intensity.

In fact, if imbalanced gains are the cause of imperfect contrast, the phase uncertainty at the WP reduces to
\begin{equation}
    \Delta \phi_\text{WP}^2 = \frac{1}{4 M n_\mathrm{min}(n_\mathrm{min} + 1)} \quad \mbox{with} \quad n_\mathrm{min} \coloneqq \min(n_0, n_0').
\end{equation}
This established result~\cite{Giese2017} highlights that in fact the lower gain of both squeezers limits the sensitivity of the result.
In fact, the additional appearance of $\sqrt{\phi_1}$ defined by \eqref{eq:phi_1} causes this change in the scaling behaviour.

Hence, operating an NLI at the WP is only favourable, if the contrast is sufficiently close to unity.
Only in this case one has a chance of observing a Heisenberg scaling of the sensitivity.
Principally, the imaging distillation technique benefits from perfect contrast as well, however, it never displays a Heisenberg scaling.
In fact, the main advantage of operating at the WP is lost when working with imperfect contrast.

\vspace{-2mm}
\subsection{Perfect Contrast Superimposed With Noise}
\vspace{-2mm}
In the remainder of this article we assume perfect contrast, \ie{} $\mathcal{C}=1$.
Now we focus on the implications of noise for the phase measurement, one of the reasons why an image distillation technique is performed.
Therefore, we choose $0 < \xi$ and $\eta<1$ in our model.
Consequently, noise enters via $n_\mathrm{n} = (1 - \eta) \langle \hat{n} \rangle$.
We compare the resulting phase uncertainties in table~\ref{tab:results_perfect_contrast}.

\begin{table}[htb]
    \caption{Comparison of phase uncertainties for perfect contrast between an MZI and NLI, as well as between the distillation technique $\displaystyle \Delta \varphi_N^2$ and an operation at the WP $\displaystyle \Delta \phi_\text{WP}^2$ including noise.
    For both interferometer types, $\eta$ describes the transmitted fraction of the interferometer signal and $M$ the number of measuring points.
    Noise enters via its detected fraction $n_\text{n}$, where we allow for a Poissonian, \ie{} $\lambda_\text{n}=0$, or thermal, \ie{} $\lambda_\text{n}=1$, noise statistics.
    While the intrinsic phase uncertainty $\Delta \varphi_0^2$ features a shot-noise scaling (MZI) and Heisenberg scaling (NLI), this behaviour cannot be exploited for the distillation technique in an NLI.
    However, an operation at the WP gives rise to uncertainties below shot noise, even in the presence of noise.
    }
\label{tab:results_perfect_contrast}
\begin{center}
    \begin{tabular}{lcc}
    \toprule
     & MZI & NLI  \\
    \midrule
    $\displaystyle \Delta \varphi_0^2$ & $\displaystyle \frac{1}{\eta n_0} \Big[1 + 2 n_\mathrm{n}+ \frac{2 n_\mathrm{n} (1+\lambda_\mathrm{n} n_\mathrm{n} )}{\eta n_0}  \Big]$ & $\displaystyle \frac{1}{4 \eta n_0 (n_0+1)} \Big[1+2n_\mathrm{n}+ \frac{n_\mathrm{n} (1 + \lambda_\mathrm{n} n_\mathrm{n})}{2\eta n_0 (1+n_0)}\Big]$\\[2em]
    $\displaystyle \Delta \varphi_N^2$ & $\displaystyle \frac{4\Delta \varphi_0^2}{M}$ & $ \displaystyle \frac{4\Delta \varphi_0^2}{M} + \frac{5}{2M}$ \\[2em]
    $\displaystyle \Delta \phi_\mathrm{WP}^2$ &%
    $\begin{array}{c}
        \displaystyle \frac{\Delta \varphi_0^2}{M} + \frac{2 \sqrt{n_\mathrm{n} (1 + \lambda_\mathrm{n}  n_\mathrm{n})}}{M (\eta n_0)^2} \\[1em]
        \times \sqrt{ n_\mathrm{n} (1 + \lambda_\mathrm{n}  n_\mathrm{n}) + \eta n_0 (1 + 2 n_\mathrm{n})}
    \end{array}$ &%
    $\begin{array}{c}
        \displaystyle \frac{\Delta \varphi_0^2}{M} + \frac{\sqrt{n_\mathrm{n} (1 + \lambda_\mathrm{n} n_\mathrm{n})}}{8 M [\eta n_0 (1 + n_0)]^2} \\[1em]
        \quad \times \sqrt{ n_\mathrm{n} (1 + \lambda_\mathrm{n} n_\mathrm{n}) + 4 \eta n_0 (1 + n_0) [1 + 2 n_\mathrm{n} + 4\eta n_0 (1+n_0)]}
    \end{array}$ \\[2em]
    \bottomrule
    \end{tabular}
\end{center}
\end{table}

\begin{figure}[h!]
    \centering
    \includegraphics[scale=1]{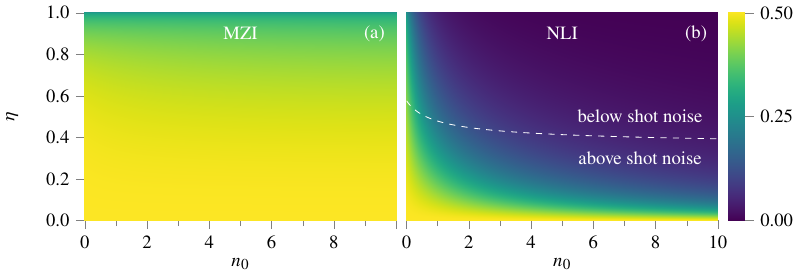}
    \caption{Advantage of operating at the WP over the distillation technique for an MZI (a) and for an NLI (b).
    We plot the ratio $\Delta \phi_\mathrm{WP}^2 / \Delta \varphi_N^2$ depending on $\eta$, describing the transmitted fraction of the interferometer signal, as well as $n_0$ corresponding to the input photon number (MZI) and gain of both squeezers (NLI).
    We assume perfect contrast, \ie{} $\mathcal{C}=1$, and choose $n_\mathrm{n} = (1-\eta) n_0$ as well as thermal noise, \ie{} $\lambda_\mathrm{n} = 1$.
    Recall that for the amplitude of the interferometer signal we have $\mathcal{A} = \eta n_0/2$ and $\mathcal{A} = 2\eta n_0(n_0+1)$ in the case of an MZI and an NLI, respectively.
    (a) In the case of an MZI, the advantage of operating at the WP over the distillation technique ranges from $1/4$ to $1/2$.
    The greatest advantage ($\Delta \phi_\mathrm{WP}^2 / \Delta \varphi_N^2=1/4$) is achieved in the limiting case of vanishing noise, \ie{} $\eta \rightarrow 1$.
    This advantage decreases for increasing noise and reaches the value $1/2$ in the limiting case.
    (b) In the case of an NLI we see that for increasing $n_0$ the phase uncertainty at the WP becomes arbitrarily better than for the distillation procedure.
    Moreover, we show by the dashed white line where the phase uncertainty at the WP corresponds to shot noise, \ie{} $\Delta \phi_\mathrm{WP}^2 = 1/n_0$.
    With that, we can identify the region in which the operation at the WP achieves a sensitivity below shot noise, which is true for any $\eta > 1/3$ in the limiting case of $n_0 \rightarrow \infty$.
    }
    \label{fig:mzi-su11-comparison}
\end{figure}

We observe that the intrinsic phase uncertainties $\Delta \varphi_0^2$ get diminished by noise for both MZI and NLI, generally degrading their scaling behaviour.
Comparing the operation of an MZI at the WP to the imaging distillation technique reveals that its phase uncertainty is no longer better by a factor of four.
Hence, the advantage of operating at the WP over the distillation technique decreases.
Turning to the NLI, we see in table~\ref{tab:results_perfect_contrast} the earlier discussed gain-independent contribution to the phase uncertainty $\Delta \varphi_N^2$ of the distillation technique, which cannot be suppressed by increasing the interrogating photon number $n_0$.
Operating at the WP enhances the sensitivity significantly, since terms introduced by noise can still be suppressed by increasing the gain $n_0$.
We visualise the ratio $\Delta \phi_\mathrm{WP}^2 / \Delta \varphi_N^2$ in figure~\ref{fig:mzi-su11-comparison} for both an MZI and NLI.
For the NLI, we additionally highlight the region in which the phase uncertainty at the WP is better than shot noise, \ie{} $\Delta \phi_\mathrm{WP}^2 \le 1/n_0$.

From figure~\ref{fig:mzi-su11-comparison}, we find the following features:
(i) Operating an MZI at the WP gives a numerical advantage over the distillation technique.
The advantage is at most a factor of $1/4$, but its exact value is determined by the fraction of noise.
For a large fraction of noise, this ratio reduces to $1/2$, so that even this advantage decreases in a noisy setup.
Needless to say, the phase uncertainty of an MZI is never below shot noise.
(ii) However, when comparing the operation of an NLI at the WP to the distillation procedure, its phase uncertainty is arbitrarily smaller for increasing gain $n_0$, even in the presence of noise.
This behaviour is expected since a noiseless NLI shows Heisenberg scaling in the high-gain regime, \ie{} $n_0 \gg 1$.
Besides this general trend, we also see for which parameter regime the phase uncertainty at the WP becomes better than shot noise, which is indicated in figure~\ref{fig:mzi-su11-comparison}\,(b) by the white dashed line where $\Delta \phi_\mathrm{WP}^2 = 1/n_0$, whose limit is $\eta=1/3$ for $n_0 \rightarrow \infty$.
In the special case of $n_\mathrm{n} = (1-\eta) n_0$ and $\eta > 1/3$, we can be better than shot noise for sufficiently large $n_0$.
Thus, our results show operating the NLI at the WP opens the possibility for sub-shot-noise phase uncertainties even in the presence of noise.

\subsection{Spontaneous Regime}
We conclude our analysis by explicitly investigating the spontaneous regime for both interferometer types, \ie{} $n_0 \ll 1$.
As before, we again set $\mathcal{C}=1$ which is in our model equivalent to $T_1 = 1/2 = T_2$ and $n_0 = n_0'$ for the MZI and NLI, respectively.
Quantum imaging distillation with two different colours has been demonstrated with an NLI in the spontaneous regime, where noise in complete analogy to our study can be introduced~\cite{Lemos2014, Fuenzalida2023, Black2023}.
Once more, we set $n_\mathrm{n} = (1-\eta) n_0$, such that the fraction of noise is tuned by $\eta$.
Table~\ref{tab:results_low_gain} summarises the resulting phase uncertainties.

\begin{table}[htb]
    \caption{Phase uncertainties in the spontaneous regime up to order $\mathcal{O}\!\left(1/M\right)$ for an MZI and NLI obtained using the distillation technique $\displaystyle \Delta \varphi_N^2$ or operating at the WP $\displaystyle \Delta \phi_\text{WP}^2$.
    We assume perfect contrast, $\eta$ describes the transmitted fraction of the interferometer signal, and $M$ the number of measuring points.
    Noise enters by setting $n_\text{n} = (1 - \eta) n_0$ such that $\eta$ determines the fraction of detected noise.
    Note that these phase uncertainties are independent of the noise statistics.}
\label{tab:results_low_gain}
\begin{center}
    \begin{tabular}{lcc}
    \toprule
    & MZI & NLI  \\
    \midrule
    $\displaystyle \Delta \varphi_N^2$ &%
    $\displaystyle 4\frac{2-\eta}{M \eta^2 n_0}$ &%
    $\displaystyle \frac{1+\eta}{2M\eta^2 n_0}$ \\[2em]
    $\displaystyle \Delta \phi_\text{WP}^2$ &%
    $\displaystyle \frac{2 + 2\sqrt{1 - \eta} - \eta}{M \eta^2 n_0}$ &%
    $\displaystyle \frac{1 + \eta + \sqrt{(1- \eta)(1 + 3 \eta)}}{8 M \eta^2 n_0 }$ \\
    \bottomrule
    \end{tabular}
\end{center}
\end{table}

Noticeably, the phase uncertainties do not depend on the noise statistics in either interferometer in the spontaneous regime and the benefit of a Heisenberg scaling ceases to exist.
Moreover, the difference between the distillation technique and operating at the WP is again only a numerical factor.
Nevertheless, it is beneficial to work at the WP due to $\Delta \phi_\mathrm{WP}^2 \le \Delta \varphi_N^2$ but there is no fundamental difference between the behaviour of the phase uncertainty.
Besides the phase uncertainties, it is important to point out that only NLIs allow for bicolour imaging contrary to MZIs.
Thus, our findings support already performed experiments in the spontaneous regime~\cite{Fuenzalida2023, Black2023}.

\section{Conclusions}
Besides allowing for bicolour imaging with undetected photons, an operation of NLIs at the WP leads to a phase sensitivity below shot noise, approaching the Heisenberg limit for perfect contrast and without noise or loss.
In our article, we investigated the impact of noise on an established phase distillation technique~\cite{Malacara2007} which has recently been used in quantum imaging distillation~\cite{Haase2022, Toepfer2022, Fuenzalida2023, Black2023} using NLIs.
We demonstrated that, not surprisingly, operating the interferometer at the WP plays a significant role in minimising the phase uncertainty resulting from the introduced noise.
For current setups operating in the spontaneous regime, we have shown that there is no fundamental difference between working at the WP or applying the distillation procedure.
However, this behaviour drastically changes when moving to the high-gain regime, where a Heisenberg-like scaling of the phase uncertainty can in principle be achieved.

In this regime of operation, the distillation technique inherently does not allow Heisenberg-like scaling, even for NLIs with perfect contrast and vanishing noise, solely caused by the thermal photon statistics of the squeezed light probing the object.
The situation is different for an operation at the WP, where we observe a phase uncertainty scaling better than shot noise, but which does not saturate a Heisenberg-type limit due to noise.
In fact, since imperfect contrast also has a deleterious effect on the sensitivity, it has to be sufficiently close to unity in order to yield a phase uncertainty below shot noise.
We furthermore show that the phase uncertainty achieved at the WP is always better compared to the distillation technique.

Even though our results demonstrated that the distillation procedure relying on equidistant phase steps uniformly distributed over the $2\pi$ interval prevents below-shot-noise sensitivities, the insights gained in our study may serve as a blueprint to nevertheless harness the potential of NLIs.
One approach could be to devise a phase-shifting algorithm for image distillation, where the device is operated in the vicinity of the WP to still achieve supersensitivity.

In perspective, it is straightforward within our formalism to include internal loss in the interferometer and extend our study for contrast imaging with objects of imperfect transmittance.
Presumably, whether a scaling of the phase uncertainty is better than shot noise depends strongly on the fraction of internal losses and the object's transmittance.
In the latter case one has to develop novel concepts for imaging in the high-gain regime, in analogy to asymmetric schemes~\cite{Kolobov2017}.

Moreover, we have only discussed a single-mode description, which might be a good approximation for imaging one pixel of an object, even though the high-gain regime is inherently multi-mode~\cite{Boyer2008, Chekhova2015, Lemieux2019}.
If a monochromatic plane wave is used as a pump field, the plane-wave modes in the squeezer decouple and are squeezed independently.
This feature allows us to view the imaging process independently per pixel.
In this case, not only perfect imaging of the output modes onto the second squeezer is necessary~\cite{Frascella2019, Scharwald2023}, but the WP may differ from pixel to pixel, which leads to further complexity.
Consequently, great flexibility is required in the second arm to adjust the WP per pixel individually.
For this purpose, spatial scans of the object or using phase-shifting algorithms or spatial light modulators for adaptive measurements~\cite{Liu2023} might be useful.
If, on the other hand, no monochromatic plane wave is used as a pump field, the modes of the non-linear process may couple, adding more complexity to a theoretical description.
In fact, the eigenmodes of squeezing are in general no plane waves, so that the entire imaging process requires a multi-mode description, which can be subtle due to time-ordering effects~\cite{Quesada2014} and the spatial profile of the pump~\cite{Scharwald2023}.
Thus, our results can also be transferred to the multi-mode scenario under appropriate conditions, although a precise multi-mode treatment is necessary as mentioned above.

\begin{acknowledgments}
We thank Jorge Fuenzalida, Sebastian T{\"o}pfer, and Sergio Adri\'{a}n Tovar P\'{e}rez for fruitful discussions.
The INTENTAS project is supported by the German Space Agency at the German Aerospace Center (Deutsche Raumfahrtagentur im Deutschen Zentrum f\"ur Luft- und Raumfahrt, DLR) with funds provided by the German Federal Ministry for Economic Affairs and Climate Action (Bundesministerium f\"ur Wirtschaft und Klimaschutz, BMWK) due to an enactment of the German Bundestag under Grant No. 50WM2177 (INTENTAS).
E.G. thanks the German Research Foundation (Deutsche Forschungsgemeinschaft, DFG) for a Mercator Fellowship within CRC 1227 (DQ-mat).
We acknowledge funding from the German Federal Ministry of Education and Research (Bundesministerium f\"ur Bildung und Forschung, BMBF) within the program ``quantum technologies – from basic research to market'' under Grant No. 13N16496 (QUANCER).
\end{acknowledgments}

\appendix
\section{Uncertainty of the Detected Signal}
\label{sect:origion_of_uncertainty}
Here we revisit the uncertainty of the detected signal \eqref{eq:variance_of_detected_signal} and provide details on the origin of the additional terms besides the variances of the interferometer signal and noise.
In the main body of the article, we introduced the photon number operators of one exit port of the interferometer as well as a noise source by $\hat{n}_\text{I}$ and $\hat{n}$, respectively.
These are related to their respective annihilation operators $\hat{a}_\text{I}$ and $\hat{a}_\text{n}$, \ie{} $\hat{n}_\text{I} =  \hat{a}^\dagger_\text{I} \hat{a}^{\phantom{\dagger}}_\text{I}$ as well as $\hat{n} =  \hat{a}^\dagger_\text{n} \hat{a}^{\phantom{\dagger}}_\text{n}$ fulfilling the bosonic commutation relations $[ \hat{a}^{\phantom{\dagger}}_i, \hat{a}^\dagger_j ] = \delta_{i,j}\hat{\mathbbm{1}}$, and $[ \hat{a}_i, \hat{a}_j ] = 0$.
The BS with transmittance $\eta \in [0, 1]$ acts as an SU(2) transformation such that the mode operator $\hat{b}$, describing the output of the BS that mixes the interferometer signal with noise is given by $\hat{b} = \sqrt{\eta} \hat{a}_\text{I} + \sqrt{1-\eta} \hat{a}_\text{n}$.
Thus, the detected photon number $N(\phi)$ is
\begin{equation}
    N(\phi) = \braket{\hat{b}^\dagger \hat{b}} = \eta \braket{\hat{n}_\text{I}} + (1-\eta)\braket{\hat{n}} + \sqrt{\eta}\sqrt{1-\eta} \braket{\hat{a}^\dagger_\text{I} \hat{a}^{\phantom{\dagger}}_\text{n} + \hat{a}^\dagger_\text{n} \hat{a}^{\phantom{\dagger}}_\text{I}} = n_\phi + n_\mathrm{n},
\end{equation}
where we set the interference term $\braket{\hat{a}^\dagger_\text{I} \hat{a}^{\phantom{\dagger}}_\text{n}}$ to zero, since we assume that the interferometer signal and noise are incoherent to each other.
Using the mode operators, we define the variance
\begin{equation}
\label{eq:eq_2_revisited}
    \Delta N^2(\phi) = \braket{ (\hat{b}^\dagger \hat{b})^2 } - \braket{\hat{b}^\dagger \hat{b}}^2 = \braket{\hat{b}^\dagger \hat{b} \hat{b}^\dagger \hat{b}} - N^2(\phi).
\end{equation}
of the detected signal.
To calculate $\braket{\hat{b}^\dagger \hat{b} \hat{b}^\dagger \hat{b}}$, the SU(2) transformation of the BS is applied, such that
\begin{subequations}
\label{eq:first_term_of_variance}
\begin{align}
    \begin{split}
    \braket{\hat{b}^\dagger \hat{b} \hat{b}^\dagger \hat{b}} &= %
    \eta^2 \braket{\hat{a}^\dagger_\text{I} \hat{a}^{\phantom{\dagger}}_\text{I} \hat{a}^\dagger_\text{I} \hat{a}^{\phantom{\dagger}}_\text{I}} %
    + (1-\eta)^2 \braket{\hat{a}^\dagger_\text{n} \hat{a}^{\phantom{\dagger}}_\text{n} \hat{a}^\dagger_\text{n} \hat{a}^{\phantom{\dagger}}_\text{n}} %
    + \eta(1-\eta) \big[%
    2\braket{\hat{a}^\dagger_\text{I} \hat{a}^{\phantom{\dagger}}_\text{I} \hat{a}^\dagger_\text{n} \hat{a}^{\phantom{\dagger}}_\text{n}} %
    + \braket{\hat{a}^\dagger_\text{I} \hat{a}^{\phantom{\dagger}}_\text{n} \hat{a}^\dagger_\text{n} \hat{a}^{\phantom{\dagger}}_\text{I}} %
    + \braket{\hat{a}^\dagger_\text{n} \hat{a}^{\phantom{\dagger}}_\text{I} \hat{a}^\dagger_\text{I} \hat{a}^{\phantom{\dagger}}_\text{n}} \big] \\
    &\phantom{=}+ \text{other contributions}
    \end{split} \\
    &= \eta^2 \braket{\hat{n}_\text{I}^2} + (1-\eta)^2 \braket{\hat{n}^2} + 2 \eta (1-\eta) \braket{\hat{n}_\text{I} \hat{n}} + \eta (1-\eta) \big[%
    \braket{\hat{n}_\text{I} (\hat{n} + \hat{\mathbbm{1}})} %
    + %
    \braket{\hat{n} (\hat{n}_\text{I} + \hat{\mathbbm{1}})} \big],
\end{align}
\end{subequations}
where the `other contributions' are of the form $\left\langle \hat{a}_\text{I}^{\dagger p} \hat{a}_\text{I}^{q} \hat{a}_\text{n}^{\dagger r} \hat{a}_\text{n}^{s} \right\rangle$ with $p, q, r, s \in \mathbb{N}_0$ and $r \neq s$.
Since we assume that the interferometer signal and noise are incoherent to each other, these `other contributions' vanish.
Using \eqref{eq:first_term_of_variance} and \eqref{eq:eq_2_revisited} together with the abbreviations introduced in the main body directly leads to \eqref{eq:variance_of_detected_signal}.

\section{Uncertainty of Tunable Phase}
\label{sect:uncertainty_of_tunable_phase}
In section~\ref{sec:phase_distl} we made the assumption that contributions $\Delta \varphi_\theta^2$, which originate in the uncertainty $\Delta \theta_j ^2$ of the tunable phase in \eqref{eq:dist_procd_uncert} and \eqref{eq:varphi_uncert}, are negligible.
In this appendix, we present the explicit expression and give some estimates for different experimental schemes of tuning the phase.
Moreover, we discuss for an NLI under which conditions such contributions are negligible. 

To express $\Delta \varphi_\theta^2$ in terms of $\Delta \theta_j ^2$ we use the relation $\partial \varphi/\partial \theta_j = -2 N(\phi_j) \cos{\phi_j }/( M \mathcal{A} \mathcal{C})$.
Moreover, it is reasonable to assume that the uncertainty of the tunable phase is equal for all measuring points, \ie{} $\Delta \theta_j = \Delta \theta$.
This relates the scanning uncertainty 
\begin{equation}
\label{eq:scan_ph_uncert}
    \Delta \varphi_\theta^2= \frac{2}{M \mathcal{C}^2} \left(1 + 3\mathcal{C}^2/4 +2 \xi +\xi^2  \right) \Delta \theta^2
\end{equation}
directly to $\Delta \theta^2$.
We see that the scanning uncertainty is independent of the noise statistics, but increases for an increasing ratio of noise to interferometer signal $\xi = n_\text{n}/\mathcal{A}$.

In the absence of noise, a contribution remains even for perfect contrast, similarly to \eqref{eq:phase_uncert_dist}.
Hence, the only possibility to suppress the scanning uncertainty is by requiring a sufficiently small $\Delta \theta^2$.
For this we give some experimental estimates, assuming a wavelength of $\SI{730}{\nano\meter}$ and consider two cases:
(i) In setups in which piezoelectric elements scan the phase, the accuracy of the adjustment can be $\SI{1}{\nano \meter}$ in the worst case.
This leads to $\Delta \theta \cong \SI{10}{\milli\radian}$.
(ii) In setups where the phase is scanned by spatial light modulators with a typical modulation depth of $4 \pi$ in $256$ steps we get $\Delta \theta \cong \SI{50}{\milli\radian}$.

These considerations are supported by already performed experiments, which have observed an overall uncertainty of $\Delta \varphi_N$ in the order of radians in the spontaneous regime~\cite{Fuenzalida2023}, which is several orders of magnitude larger than the limit given by $\Delta \varphi_\theta^2$.
Of course, this observation is a consequence of the operation in the spontaneous regime, \ie{} $n_0 \ll 1$.
Table~\ref{tab:results_low_gain} shows that for both interferometer types $\Delta \varphi_N^2$ scales with $1/n_0$, in contrast to the scanning phase \eqref{eq:scan_ph_uncert} that is independent of $n_0$ or $\mathcal{A}$ to lowest order.
Thus, it is reasonable to assume that this term is the leading order contribution of the overall phase uncertainty and $\Delta \varphi_\theta^2$ as well as $\Delta \theta^2$ are negligible.
In fact, in the spontaneous regime we expect no limitation of the achievable sensitivity that arises from the uncertainty of the tunable phase.

For quantum imaging distillation in the high-gain regime, the phase sensitivity is limited by the gain-independent contribution $2 (1+ \mathcal{C}^2/4)/ (M \mathcal{C}^2)$, so that it is realistic that \eqref{eq:scan_ph_uncert} is smaller, especially since both show the same scaling in $M$.
Hence, we assume that the uncertainty of the tunable phase is no limiting factor.

However, performing a phase measurement at the WP has the intrinsic benefit of exhibiting a sub-shot-noise sensitivity.
In this case, we find the requirement that $\Delta \varphi_\theta^2$ is smaller than the targeted sensitivity.
In principle, the uncertainty $\Delta \varphi_\theta^2$ can be minimised by independent measurements.
But to find the WP with minimal phase uncertainty, one has to characterise the signal by independent measurements prior to the experiment.
In fact, not only $\theta$, but also $\mathcal{A}$ and $\mathcal{C}$ have to be determined with sufficient accuracy to find the WP.
For a fair comparison, the required time and resources for a characterisation must be included in the total error budget.
Often, though, this time is omitted in purely interferometric setups, which use continuous operation, since it amounts to only a small fraction of the total averaging time.
Whether it is necessary to include the characterisation time in imaging experiments depends on the specific setup, the required phase uncertainties, and the overall averaging times.
However, for imaging applications with frequently changing phase objects, such an assessment ought to be included in the overall resources.

\bibliography{literature.bib}

\end{document}